\documentclass{agujournal2018}
\usepackage{apacite}
\usepackage{url} 
\usepackage{float}
\usepackage{natbib}
\usepackage{amsmath}
\usepackage[english]{babel}

\draftfalse

\journalname{Earth and Space Science}

\begin{document}

%
%


\title{Middle-atmosphere dynamics observed with
  a portable muon detector}

%
%



\authors{M. Tramontini\affil{1,2}, M. Rosas-Carbajal\affil{3},
  C. Nussbaum\affil{4}, D. Gibert\affil{5}, J. Marteau\affil{1}}

\affiliation{1}{Institut de Physique Nucl\'eaire de Lyon, UMR 5822, CNRS-IN2P3, Universit\'e de Lyon, Universit\'e Claude Bernard Lyon 1, France}
\affiliation{2}{CONICET - Facultad de Ciencias Astron\'omicas y Geof\'isicas, Universidad Nacional de La Plata, Argentina}
\affiliation{3}{Universit\'e de Paris, Institut de Physique du Globe
  de Paris, CNRS, UMR 7154, F-75238 Paris,
  France}
\affiliation{4}{Swiss Geological Survey at swisstopo, Seftigenstrasse 264, CH-3084 Wabern, Switzerland}
\affiliation{5}{Univ. Rennes, CNRS, G\'eosciences Rennes, UMR 6118, F-35000, Rennes, France}





\correspondingauthor{Marina Rosas-Carbajal}{rosas@ipgp.fr}




\begin{keypoints}
\item We report muon rate variations associated to temperature changes
  in the middle atmosphere observed with a portable muon detector 
\item The effect is significant both for seasonal and short-term temperature variations, even under low-opacity conditions at mid-latitudes
\item We highlight potential applications on atmosphere dynamics
 and the need to account for these phenomena in
  geophysical applications  \\

\end{keypoints}

%
%


\begin{abstract}
\justifying
In the past years, large particle-physics experiments have shown
that muon rate variations detected in underground laboratories are
sensitive to regional,
middle-atmosphere temperature variations. Potential applications
include tracking short-term atmosphere dynamics, such as Sudden
Stratospheric Warmings. We report here that such
sensitivity is not only limited to large surface detectors under
high-opacity conditions. We use a portable muon detector conceived for
muon tomography for geophysical applications and we
study muon rate variations observed over one year of measurements at
the Mont Terri
Underground Rock Laboratory, Switzerland (opacity of $\sim 700$ meter water equivalent).
We observe a direct correlation between middle-atmosphere
seasonal temperature variations and muon rate. Muon rate variations
are also sensitive to the abnormal atmosphere
heating in January-February 2017, associated to a Sudden Stratospheric
Warming. 
Estimates of the effective temperature coefficient for our
particular case agree with theoretical models and with those
calculated from 
large neutrino experiments under comparable conditions.
Thus, portable muon detectors may be useful to 1) study seasonal and short-term middle
atmosphere dynamics, especially in locations where data is lacking such
as mid-latitudes; and 2) improve the calibration of the effective
temperature coefficient for different opacity conditions. 
Furthermore, we highlight the importance of assessing the impact of
temperature on muon rate variations when considering geophysical
applications. Depending on latitude and opacity conditions, this
effect may be large enough to hide subsurface density variations due
to changes in groundwater content, and should therefore
be removed from the time-series.

\end{abstract}

%
%

%


%
%
%
%

\section{Introduction}
\justifying

First observed in 1952 using radiosonde measurements
\citep{scherhag52}, Sudden Strato\-spheric Warmings (SSWs) are
extreme wintertime circulation anomalies that 
produce a rapid rise in temperature
in the mid to upper polar stratosphere (30-50 km).
SSW effects on middle-atmosphere dynamics have lifetimes of approximately 80 days \citep{limpasuvan2004life}.
They are the
clearest and strongest manifestation of dynamic coupling
throughout the whole atmosphere-ocean system \citep{o2014effects,goncharenko2010unexpected,liu2002study}.
Following a major SSW, the high altitude winds reverse to flow westward
instead of their usual eastward direction. 
This reversal often results in dramatic surface temperature reductions
in mid-latitudes, particularly in Europe, which suggests the
possibility of monitoring the stratosphere for predicting extreme
tropospheric weather \citep{thompson2002stratospheric}. 
The frequency of SSWs may increase due to
global warming \citep{schimanke2013variability,kang2017more}. 
While many studies have focused on the characterization of SSWs through
observation and modeling dynamics at high latitude
regions, observation studies at mid-latitudes are rare and could be
crucial to better understand the phenomena 
\citep{yuan2012wind,sox2016connection}.

Cosmic muons represent the largest proportion of 
charged particles reaching the surface of the Earth, yielding a flux
of $\sim$ 70 m${}^{-2}$s${}^{-1}$sr${}^{-1}$ for particles above 1
GeV  \citep{PDG18}. They are a product of the
primary cosmic rays interaction with the atmosphere, 
which produces short-lived
mesons, in particular, charged pions and kaons. These particles decay
into muons that easily penetrate the atmosphere and may reach the
surface of the Earth.
The flux of muons decreases as muons travel through an increasing amount
of matter. Thus, only the most energetic muons can reach
underground detectors \citep{gaisser2016cosmic}.  
The muon production process requires that the parent mesons did not
undergo destructive interactions with the propagating medium before they decay
\citep{grashorn2010atmospheric}.
Thus, changes in the atmospheric properties, in particular
in its density, may have large impacts on the muon flux measured at
ground level, either by affecting the parent mesons survival probabilities before decay
or by affecting the rate of absorption of the muons themselves along
their path down from their production level.

An increase in the atmospheric temperature lowers the atmospheric
density. Temperature changes in the atmosphere may therefore affect the
production of muons \citep{gaisser2016cosmic}.
The decrease in atmospheric density increases the mean free path of the mesons and therefore
their decay probability, thus increasing the muon flux.
The effect is more important for high-energy muons, which result from
high-energy mesons with larger lifetime due to time dilation and
therefore with longer paths in the atmosphere.
This increases their interaction probability before decay
\citep{grashorn2010atmospheric},
 thus one expects high-energy muons to be more sensitive to
temperature changes.
The opacity is the integrated density along a travel path. It is 
used to quantify the amount of matter encountered by the muons 
and is generally expressed in meter water equivalent (mwe).
Detectors in high-opacity conditions are
more likely to register the effects of temperature variations in the
atmosphere.
Notice that the low-energy muons may also be affected by 
temperature
changes because their own interaction probability with
the atmosphere along their path down to the Earth depends on the
atmospheric density. Indeed, this effect has been observed in low opacity
conditions \citep[e.g.][]{jourde2016monitoring}, but is not relevant
for detectors deeper than 50 mwe \citep{ambrosio97}.
The variations in the cosmic
muon flux caused by atmospheric temperature changes can be
treated in terms of an effective temperature
\citep{barrett52,ambrosio97}.
This effective temperature is a weighted average of the
atmosphere's temperature profile, with weights related to the
altitudes where muons are produced \citep{grashorn2010atmospheric}. 

Modulation of the cosmic muon flux produced by seasonal variations in
the atmospheric temperature have been reported for large 
detectors (AMANDA: \citet{bouchta1999seasonal}, Borexino:
\citet{agostini2019modulations}, Daya Bay: \citet{an2018seasonal},
Double Chooz: \citet{abrahao2017cosmic}, GERDA: \citet{GERDA16}, IceCube:
\citet{desiati2011seasonal}, LVD: \citet{vigorito2017underground},
MACRO: \citet{ambrosio97}, MINOS:
\citet{adamson2014observation,adamson2010observation}, OPERA:
\citet{agafonova2018measurement}). 
\citet{osprey09} and \citet{agostini2019modulations} also report that
measured muon rates
are sensitive to short-term variations (day scale) in the thermal state
of the atmosphere, such as the occurrence of SSWs. 
\cite{agafonova2018measurement} observed short-term, non-seasonal
variations in latitudes as low as 42$^\circ$ N, in Italy.

The previously mentioned studies highlight the potential of muon
measurements to characterize and monitor middle atmosphere
dynamics. However, all these studies were conducted by large-scale, general-purpose particle 
detectors, specifically built for neutrino and high-energy particle
experiments.
Most of them were placed hundreds of meters
under\-ground, which improves data sensitivity to atmospheric
effects by filtering out low-energy muons.
The detection surface of these systems are huge compared to 
portable ones, which are used for geoscience applications such as
characterizing the density structure of volcanoes \citep[e.g.][]{rosas2017three}.
Recently, muon rate variations following the passage of a thundercloud were
reported by \citet{hariharan2019measurement} using a relatively 
large detector (6$\times 6$$\times 2$ m${}^3$).  
To the best of our knowledge, no experiment has reported the
sensitivity of portable muon detectors to middle atmosphere dynamics,
especially under relatively low opacity conditions.

In this paper, we study seasonal and short-term variations
in the muon rate observed with a portable muon detector installed at
the Mont Terri Underground Rock Laboratory (Switzerland, 47.4$^\circ$
N). We first present our detector and the general conditions under
which the measurements were taken. We then analyze the variations
observed and compare them to atmospheric temperature and
middle-atmosphere dynamics data. Finally, we discuss the
implications of our observations both for the atmospheric science and geophysics
communities, the latter aiming to characterize density
variations in the subsurface with muon data.

\section{The muon detector}

\begin{figure}[H]
	\centering
	\includegraphics[width=1\linewidth]{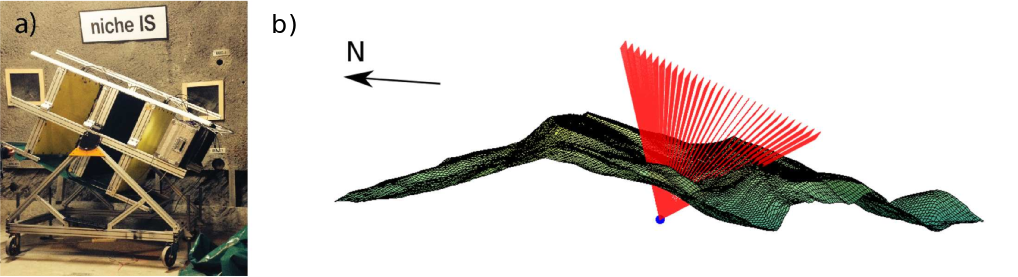}
	\caption{a) The muon telescope deployed in the Mont Terri URL. b) Telescope's position (blue) and axes of observation (red), along with the topography.}
	\label{fig:telescope}
\end{figure}

Our portable muon detector was conceived
for geoscience applications by the DIAPHANE project
\citep[e.g.,][]{marteau2012muons,marteau2017diaphane}. 
It is equipped with 3 plastic scintillator matrices of 80 cm width
composed by $N_x=N_y=16$ scintillators bars, in the horizontal and
vertical directions,
whose interceptions define $16 \times 16$ pixels of $5 \times 5$
cm${}^2$. 
When a muon passes through the 3 matrices (i.e., an
``event'' is registered),
3 hits are recorded in time coincidence, with a resolution better than
1 ns \citep{marteau2014telescope}, enabling us to
reconstruct its trajectory from the sets of pixels fired in each matrix.
We apply a selection based on the goodness of the
reconstructed trajectory in order to filter out random coincidences, i.e, three 
coincident fired pixels that do not align.
If the reconstructed trajectories using
two consecutive matrices differ by more than one pixel, 
in either the horizontal or the vertical direction, the event is discarded.
More details on the hit selection and the technique applied
to determine the propagation directions of muons through the detector matrices 
can be found 
in \citet{jourde2015thesis} and in \citet{marteau2014telescope}.
The distance between the front and rear matrices is set to 100 cm for
this study (Fig. \ref{fig:telescope}a). 
Because of the large volume of rock studied compared to the detector
size, we admit a point-like approximation of the detector \citep{lesparre2010geophysical}. 
With this approximation, given that two points are sufficient to uniquely determine
a direction, events whose pair of pixels in the front and the rear
matrices share the same
relative direction are
considered to correspond to the same trajectory. 
This yields a total of
$(2N_x-1) \times (2N_y-1) = 961$ axes of observation studied
(represented in 
Fig. \ref{fig:telescope}b). 

The passage of muons is detected with
wave-length shifting optical fibers that transport the photons
generated by the scintillators to the photomultiplier, where they are
detected based on a time coincidence logic. 
The optoelectronic chain has been developed from high-energy particle experiments on the concept of the autonomous,
Ethernet-capable, low power, smart sensors
\citep{marteau2014telescope}. 
In order to support strenuous field conditions, besides being sensitive
the detector is also robust, modular and transportable
\citep{lesparre2012design}.
In this experiment, the muon detector was deployed in the Mont Terri
Underground Rock Laboratory (URL) and acquired data
for 382 days between October 2016 and February 2018.
The minimum and the maximum amount of rock traversed by muons registered by the detector
are of approximately 200 and 500 m, respectively.
Prior to the underground measurements, a calibration
experiment was performed by measuring the open-sky muon flux at the
zenith, from which we register a total acceptance of 1385 cm${}^2$ sr
for our data set \citep{lesparre2010geophysical}.

\section{Methodology}

Our data set consists of a list of muon detections called ``events''. 
Each event is characterized by the arrival time and the direction of
the particle (possible directions shown in Fig. \ref{fig:telescope}b).
From these data, we compute the average cosmic muon rate, $R$, using a
30-day width Hamming moving average window \citep{hamming1998digital}. In order to increase the
signal to noise ratio and, therefore, to improve the statistics in
our analysis, we merge the signals from all the 
directions together \citep[e.g.][]{jourde2016muon}. Such a merging is done exclusively
to compute $R$.

Seasonal variations in $R$, caused by the temperature changes in the atmosphere, can be treated in terms of an effective temperature \citep{barrett52}, $T_{\text{eff}}$:
\begin{equation}
\dfrac{\Delta R}{\left\langle R \right\rangle } = \alpha_\text{T} \dfrac{\Delta T_{\text{eff}}}{\left\langle T_{\text{eff}} \right\rangle} \ ,
\label{eq:relationship}
\end{equation}
\noindent where $\alpha_\text{T}$ is the effective temperature coefficient, $\left\langle R \right\rangle$ is the mean muon rate and $\left\langle T_{\text{eff}} \right\rangle$ is the mean effective temperature.
$T_{\text{eff}}$ is defined as the temperature of an isothermal atmosphere that produces the same
meson intensities as the actual atmosphere. Thus, it is related to the atmosphere's temperature profile, 
and it is associated to the altitudes where observed muons are produced. We use the parametrization
given by \citet{grashorn2010atmospheric}:

\begin{eqnarray}
T_{\text{eff}}  =  \dfrac{\int_{0}^{\infty} W(X) T(X) dX}{\int_{0}^{\infty} W(X) dX} \ ,
\label{eq:Teff}
\end{eqnarray}

\noindent where the temperature, $T(X)$, is measured
as a function of atmospheric depth, $X$.
The weights, $W(X)$, contain the contribution of each atmospheric depth to the
overall muon production. These weights depend on the threshold energy
$E_{\text{th}}$, that is,
the minimum energy required for a muon to survive a
particular opacity in order to reach the underground detector.
Since $T(X)$ is measured at discrete levels of $X$, we perform a numerical
integration based on a quadratic interpolation
between temperature measurements to obtain $T_{\text{eff}}$. 

The effective temperature will be different for different zenith angles. To compare $T_\text{eff}$ variations to our measured muon rates, we need to account for this dependence.
Following \citet{adamson2014observation}, we bin the zenith angle distribution and
calculate a weighted effective temperature, $T_\text{eff}^\text{weight}$, as:

\begin{equation}
T_\text{eff}^\text{weight} = \sum_{i=1}^{M} F_i \cdot T_\text{eff}(\theta_i) \ ,
\end{equation}

\noindent where $M$ is the number of zenith-angle bins, $T_\text{eff}(\theta_i)$
is the effective temperature in bin $i$ and $F_i$ is the fraction of muons
observed in that bin. The formula for $T_\text{eff}(\theta_i)$ is similar to Eq. (\ref{eq:Teff}), but the atmospheric 
depth is replaced by $X/\cos\theta$
and $E_{\text{th}}$ is calculated for each zenith angle as
well.
From now on, we will refer to $T_\text{eff}^\text{weight}$ as
$T_\text{eff}$. These values are calculated four times a day and then
day-averaged, and the resulting
standard deviation is used
as an uncertainty estimate of the effective temperature daily mean value.
Thus, a representative value of effective temperature is calculated for each day,
which fully accounts for the particular setup of our experiment.

The goodness of fit of the linear relationship in  Eq. (\ref{eq:relationship}) can be
quantified by the Pearson correlation coefficient $r$.
This parameter is equal to $\pm 1$ for a full
positive/negative linear correlation, respectively, and 0 for no
correlation. 
We perform a linear regression between the relative muon rate and
effective temperature variations using Monte Carlo simulations.
In this way, we can account for error bars in both
variables and compute the uncertainty of the fitted parameters. 
Following \citet{adamson2010observation}, the intercept is fixed at
zero and the slope of the linear fit is the effective temperature
coeffi\-cient, $\alpha_\text{T}$. 
To evaluate the effects of systematic uncertainties we modify $\left\langle T_{\text{eff}} \right\rangle$ and the
parameters involved in the computation of 
$T_{\text{eff}}$ \citep[i.e. the twelve input parameters in $W(X)$ ,
c.f.][]{adamson2010observation} 
and recalculate the effective temperature
coefficient, $\alpha_{\text{T}}$.
These systematic errors are added in quadrature to the statis\-tical
error obtained from the linear fit in orden to obtain the experimental
value of $\alpha_{\text{T}}$. 

We also use Monte Carlo simulations to determine the theoretical expected
value of the effective temperature coefficient,
$\alpha_\text{T}^{\text{theory}}$, in order to compare it with the
experimental one. 
Muon energy, $E_\mu$, and zenithal angle, $\theta$, are
randomly sampled from the differential muon spectrum given by
\citet{gaisser2016cosmic} and corrected for altitude according to
\citet{hebbeker2002compilation}. 
Then, the muon is randomly assigned an azimuthal angle, $\phi$, according
to a uniform probability distribution. 
The overburden opacity in the Mont
Terri URL is determined for each combination of  ($\phi$,
$\theta$) from our muon data set, together with the
corresponding $E_{\text{th}}$ \citep{PDG18}. 
We continue the Monte Carlo sampling until 
we obtain 10,000 successful events that satisfy $E_\mu >
E_{\text{th}}$, for which we compute the $\alpha_\text{T}^{\text{theory}}$
distribution using the expression derived by
\citet{grashorn2010atmospheric}.
Next, we determine the value of $\alpha_\text{T}^{\text{theory}}$ and
its uncertainty as the mean and standard deviation of the
distribution, respectively. The systematic uncertainty is the one
reported by \citet{adamson2014observation}.

We look for the ocurrence of SSWs during the acquisition period
using the definition of a major SSW given by \citet{charlton2007}. 
A major mid-winter warming is considered to occur when the
zonal mean zonal wind at 60$^\circ$N and 10 hPa become easterly during
winter. The first day on which this condition
is met is defined as the central date of the warming.
The zonal mean zonal wind is the average east-west (zonal) 
wind speed along a latitude circle.
To ensure that only major mid-winter warmings are identified, 
cases where the zonal mean zonal wind does not reverse back to westerly for 
at least 2 weeks prior
to their seasonal reversal to easterly in spring are assumed to be final warmings, 
and as such are discarded.
SSWs typically manifest as a displacement or a splitting of the polar vortex
\citep{charlton2007}, a cyclone residing on both of the Earth's
poles that goes from the mid-troposphere into the stratosphere.

\section{Results}\label{sec:Results}

\begin{figure}[H]
	\centering
	\includegraphics[width=1\linewidth]{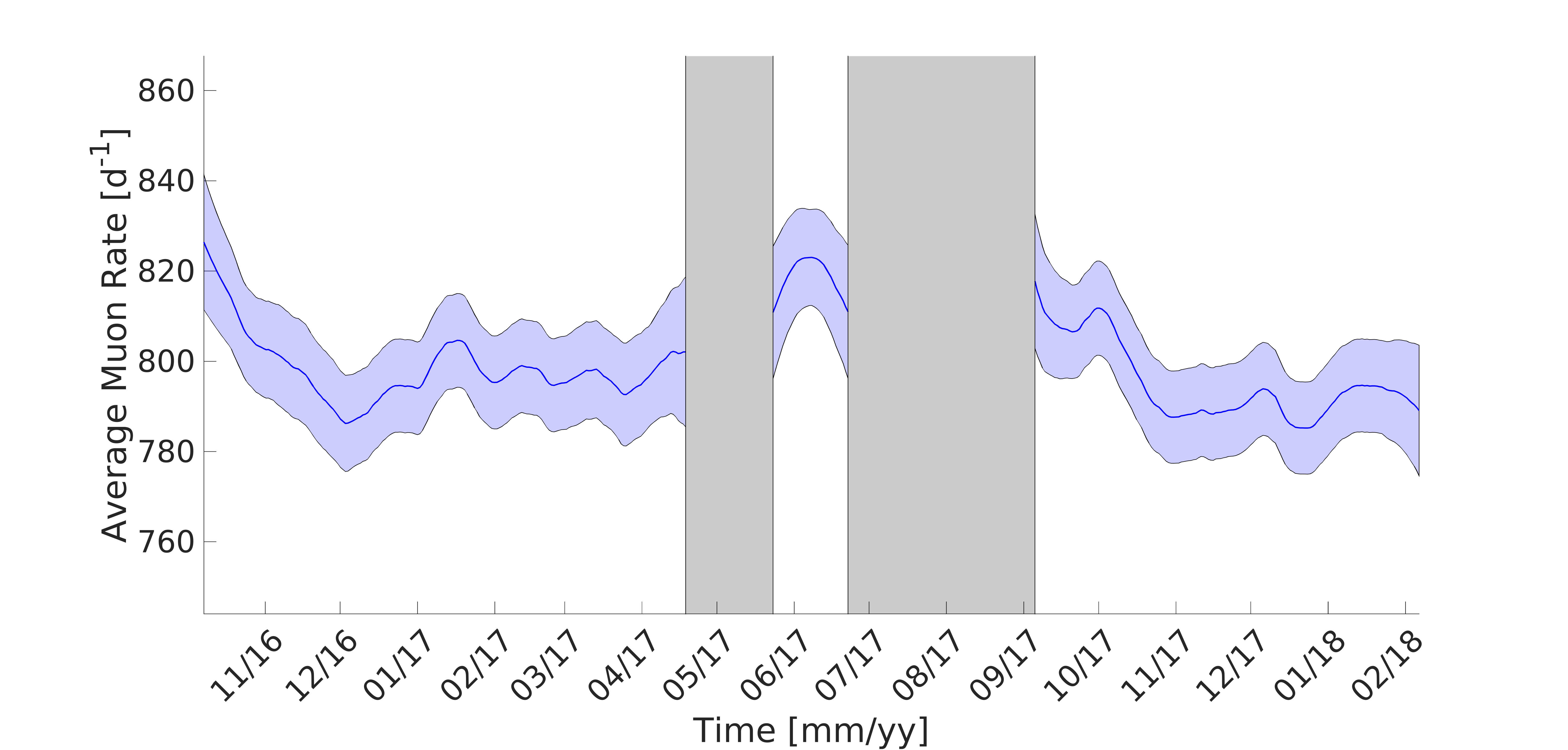}
	\caption{Average cosmic muon rate as a function of time, computed
		using a 30-day width Hamming moving average window. The colored
		surface delimits the 95\% confidence interval. Gray bars
		indicate periods where the acquisition was interrupted for
		work
		in the Mont Terri URL.}
	\label{fig:absoluteFlux}
\end{figure}

Based on 382 days of data, the average daily rate of cosmic muons in
the Mont Terri URL is of $(800 \pm 10)$ d$^{-1}$, calculated by counting all the muons 
detected each day no matter their direction or the altitude 
at which they were produced.
We also compute an average muon rate for 
each axis of observation, which we use to estimate the corresponding opacity values. Minimum and maximum opacities are of approximately 500 and 1500 mwe, respectively, while the average opacity considering all possible directions is of $(700 \pm 160)$ mwe. 
The cosmic muon rate
presents significant variations in time (Fig. \ref{fig:absoluteFlux}).
Maximum rate values occur close to the summer periods while minimum
rate values occur during winter times. 

\begin{figure}[H]
	\centering
	\includegraphics[width=1\linewidth]{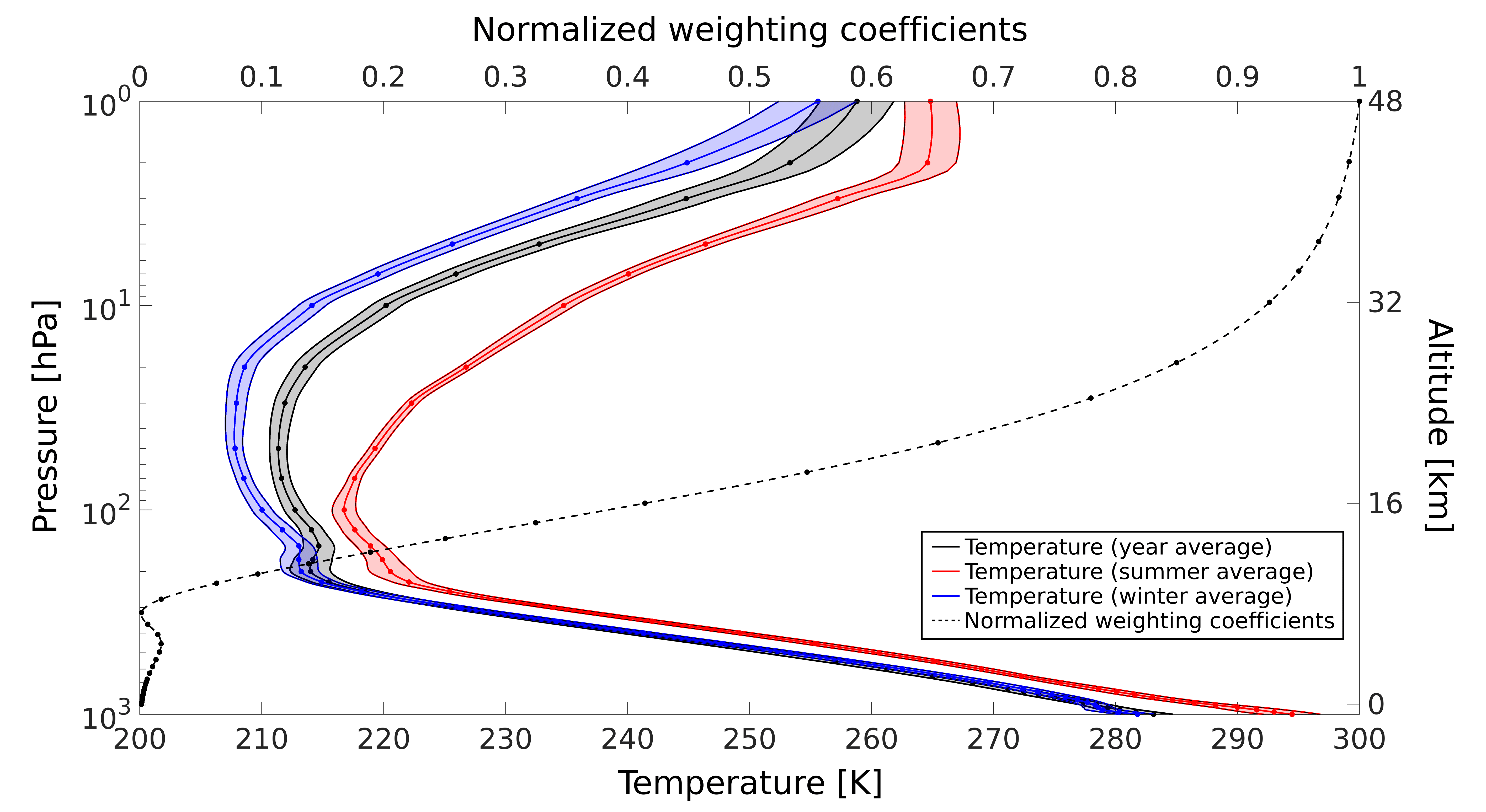}
	\caption{Atmospheric temperature profiles (solid lines) above the Mont Terri site, and weighting coefficients (dashed line) used to calculate $T_\text{eff}$, as a function of pressure level and altitude. The dots represent the 37 pressure levels for which the temperature data sets are provided by the ECMWF. The right vertical axis represents approximate altitudes corresponding to the pressure levels on the left vertical axis. The summer average temperature (solid red line) and the winter average temperature (solid blue line) are computed considering a period of 1.5 months in each season during 2017. The colored surfaces represent the $\pm 1$ standard deviation in each curve. The effective temperatures of each profile are: \mbox{$T_\text{eff}^{\text{year}}=(217\pm1)$ K}, $T_\text{eff}^{\text{summer}}=(225\pm1)$ K and $T_\text{eff}^{\text{winter}}=(214\pm1)$ K.
	}
	\label{fig:TandW}
\end{figure}

We use the ERA5 data set offered by the European Centre for
Medium-range Weather Forecast (ECMWF), which is a climate reanalysis
data set produced using 4D-Var data assimilation
\citep{era5}. Temperature data consist of interpolated
(${0.25}^{\circ}$ by ${0.25}^{\circ}$) globally gridded data on 37
atmospheric pressure levels from 0 to 1000 hPa, listed four times a
day (00:00 h, 06:00 h, 12:00 h and 18:00 h). From this data set, we
interpolate the temperature profiles at Mont Terri URL location.
In Fig. \ref{fig:TandW} we present the typical atmospheric temperature profiles at Mont Terri for summer, winter and a year average over the analysis period. We also display in the same plot the corresponding normalized weighting coefficients 
$W$ as a function of pressure levels, used to compute $T_\text{eff}$.
The largest temperature changes occur above $\sim$16 km, where the weighting coefficients are more significant. The effective temperatures corresponding to the average curves and $\theta = 0^{\circ}$ are given by
$T_\text{eff}^{\text{year}}=(217\pm1)$ K, $T_\text{eff}^{\text{summer}}=(225\pm1)$ K and $T_\text{eff}^{\text{winter}}=(214\pm1)$ K. There is thus a difference of $\sim$10 K between typical summer and winter conditions.  

\begin{figure}[H]
    \centering
    \includegraphics[width=1\linewidth]{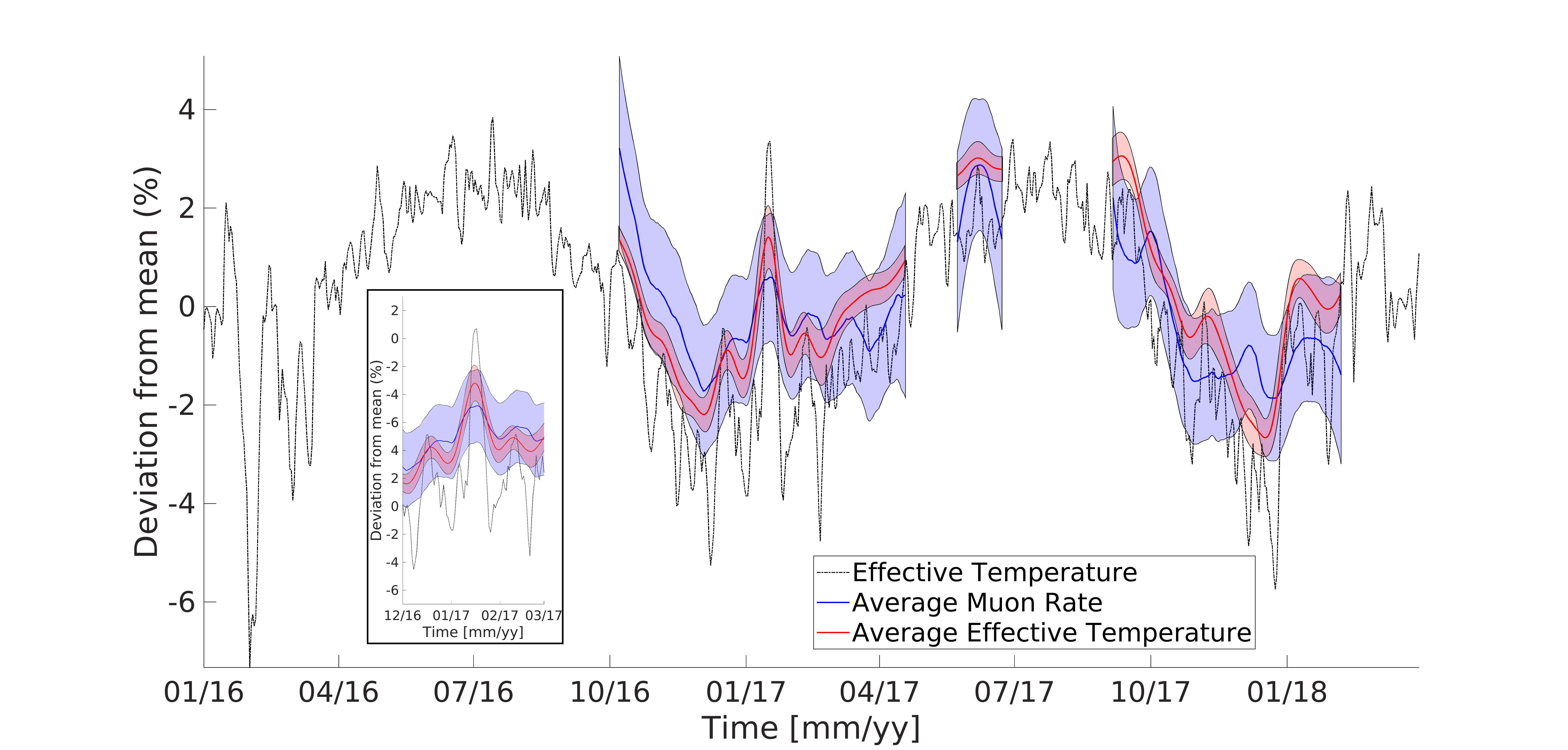}
    \caption{Daily percent deviations from the mean of the average
      cosmic muon rate, the daily effective temperature, and the
      average effective temperature computed using a 30 days width
      Hamming moving average window. The colored surfaces delimit the 95\% confidence
      interval associated to each curve. The inset displays a zoom around the period of time in which a major SSW is detected.}
    \label{fig:dfm}
\end{figure}

\begin{figure}[H]
    \centering
    \includegraphics[width=1\linewidth]{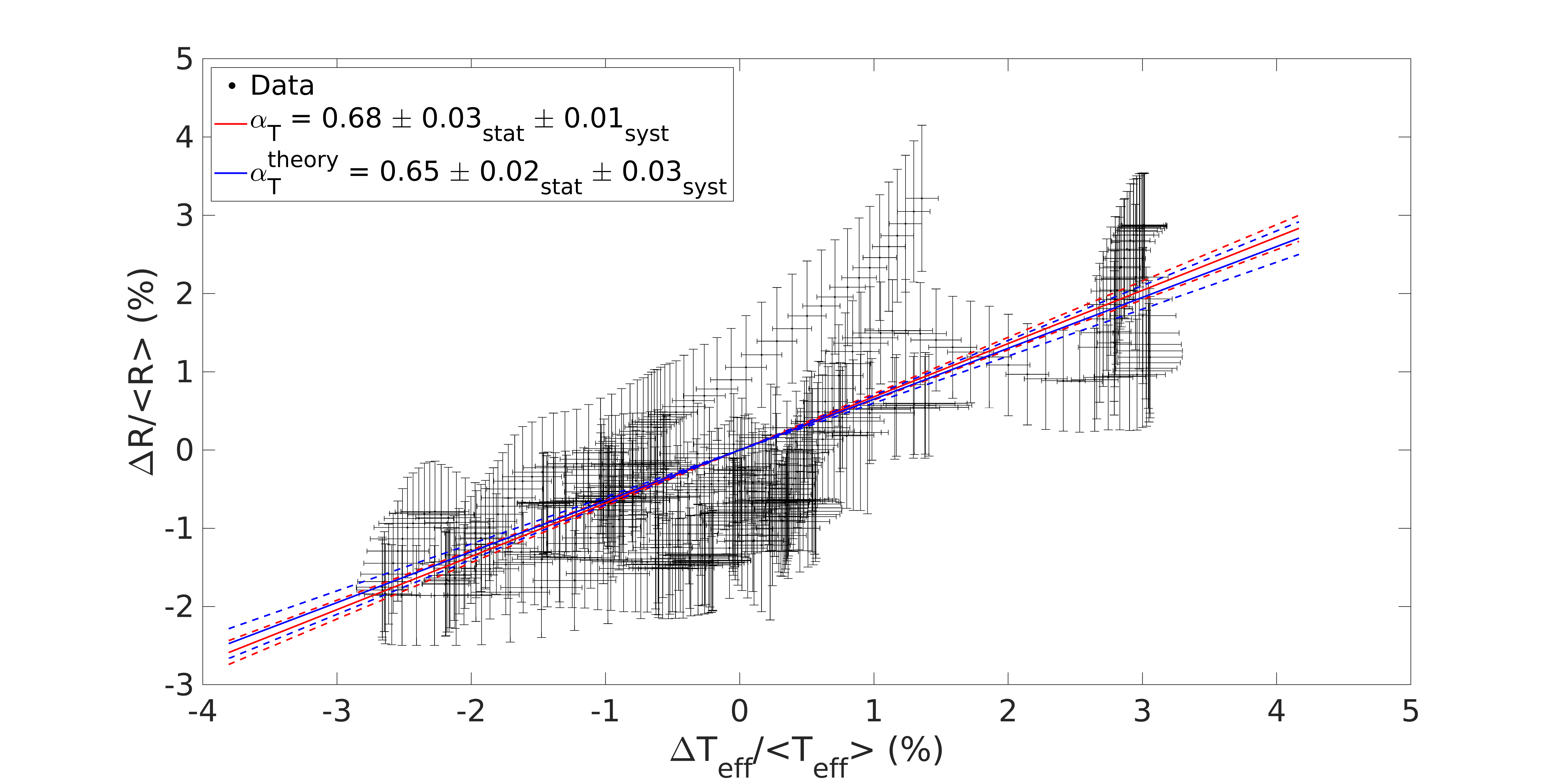}
    \caption{Average cosmic muon rate relative variation versus
      average effective temperature relative variation, fitted with a
      line with the \textit{y}-intercept fixed at 0. The resulting slope is $\alpha_T = 0.68 \pm 0.03_{\text{stat}} \pm 0.01_{\text{syst}}$ and is represented with a red line. The blue line represents the theoretical expected value of $\alpha_{\text{T}}^{theory} = 0.65 \pm 0.02_{stat} \pm 0.03_{syst}$. The dotted lines represent the uncertainty of each one of the values.}
    \label{fig:alpha}
\end{figure}

We compare the variations in the muon rate to the variations 
in the effective temperature in Fig. \ref{fig:dfm} 
in terms of relative variations (see Eq. \ref{eq:relationship}).
For consistency, we also apply a Hamming moving average window
of 30 days to the $T_\text{eff}$ time series.
The two average curves evolve similarly in time.
Indeed, the Pearson correlation coefficient between the
deviation from mean of the average muon rate and that of average
effective temperature yield a
value of 0.81.  
We compute a linear fit between the two data sets (see Methodology), which yields
an effective temperature coefficient of $\alpha_{\text{T}} = 0.68 \pm
0.03_{stat} \pm 0.01_{syst}$, with $\chi^2/\text{NDF} = 414/381$
being the reduced $\chi^2$ of the fit (Fig. \ref{fig:alpha}). 
The largest contribution to the systematic error in $\alpha_{\text{T}}$
comes from the $\pm 0.06$ uncertainty in the meson production ratio
\citep{barr2006uncertainties}, the $\pm 0.31$ K uncertainty in the
mean effective temperature \citep{adamson2010observation} and the $\pm
0.026$ TeV uncertainty in $E_\text{th}$,
which results from the distribution of opacities along the axes of
observation.  
To discard possible systematic biases, we also performed a linear fit allowing for a non-zero y intercept. The fit resulted in an estimated value of zero within one standard deviation uncertainty for this intercept, and a slightly lower value of $\alpha_{\text{T}} = 0.67 \pm
0.03_{stat} \pm 0.01_{syst}$ for the effective temperature coefficient.

\begin{figure}[H]
    \centering
    \includegraphics[width=.8\linewidth]{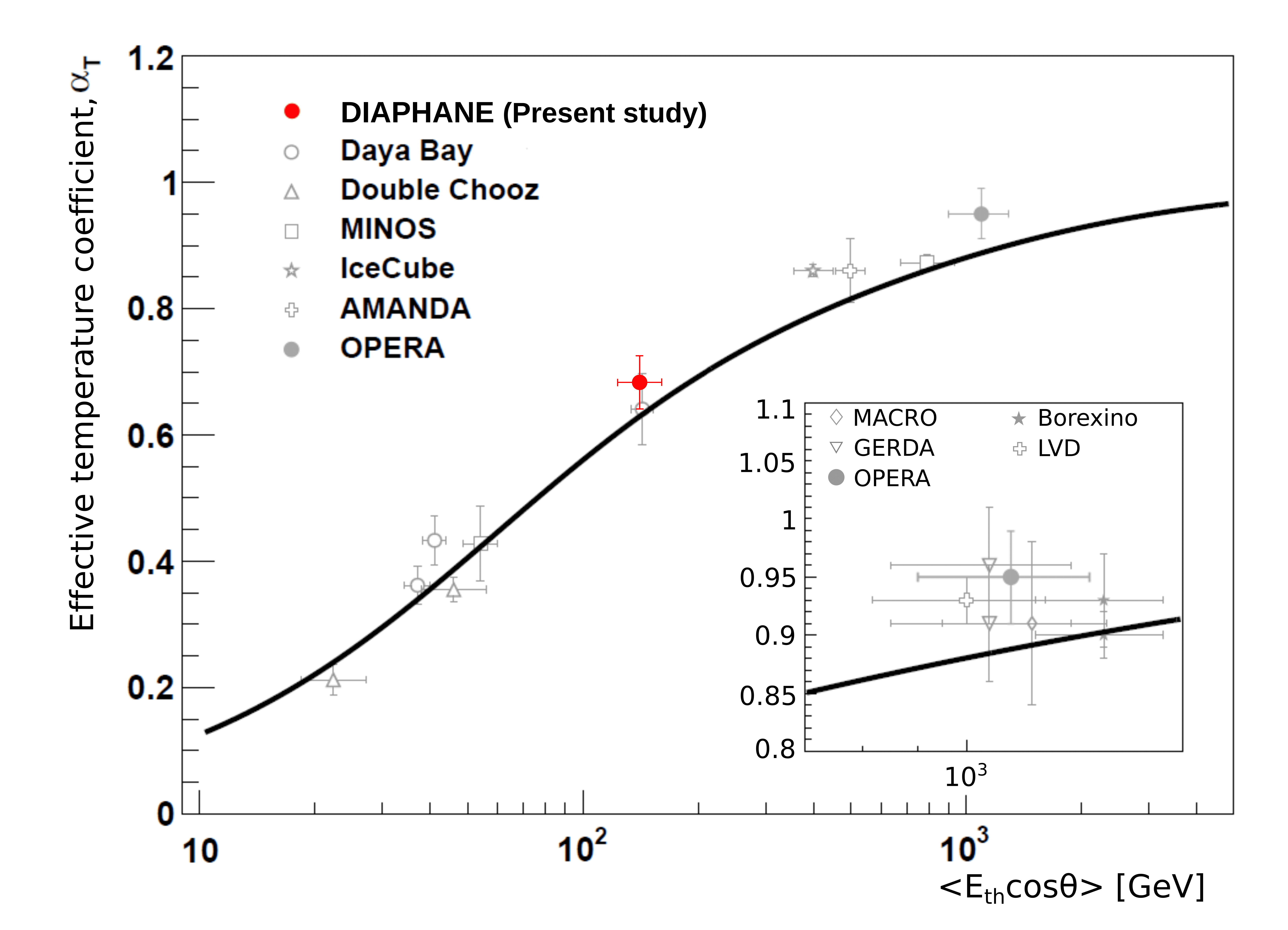}
    \caption{Experimental values of the effective temperature
      coefficient as a function of \mbox{$\left\langle E_\text{th}
      \cos\theta \right\rangle$}. The red dot represents the present study. The continuous black line represents a
      theoretical model. The insert plot show the experiments
      performed at the underground Gran Sasso Laboratory. 
      Figure adapted from \citet{agafonova2018measurement}}
    \label{fig:expValues}
\end{figure}

The theoretical expected value was found to be
$\alpha_{\text{T}}^{theory} = 0.65 \pm 0.02_{stat} \pm 0.03_{syst}$.
Thus, the experimentally estimated value is consistent with the theoretical one
within one standard deviation.
In Fig. \ref{fig:expValues} we present our estimated value of $\alpha_{\text{T}}$ along with
a theoretical model accounting for pions and kaons \citep{agafonova2018measurement},
and estimates from other experiments. Our estimate 
is consistent with the one obtained by \citet{an2018seasonal} in
similar opacity conditions, and with the theoretical model.

\begin{figure}[H]
	\includegraphics[width=1\textwidth]{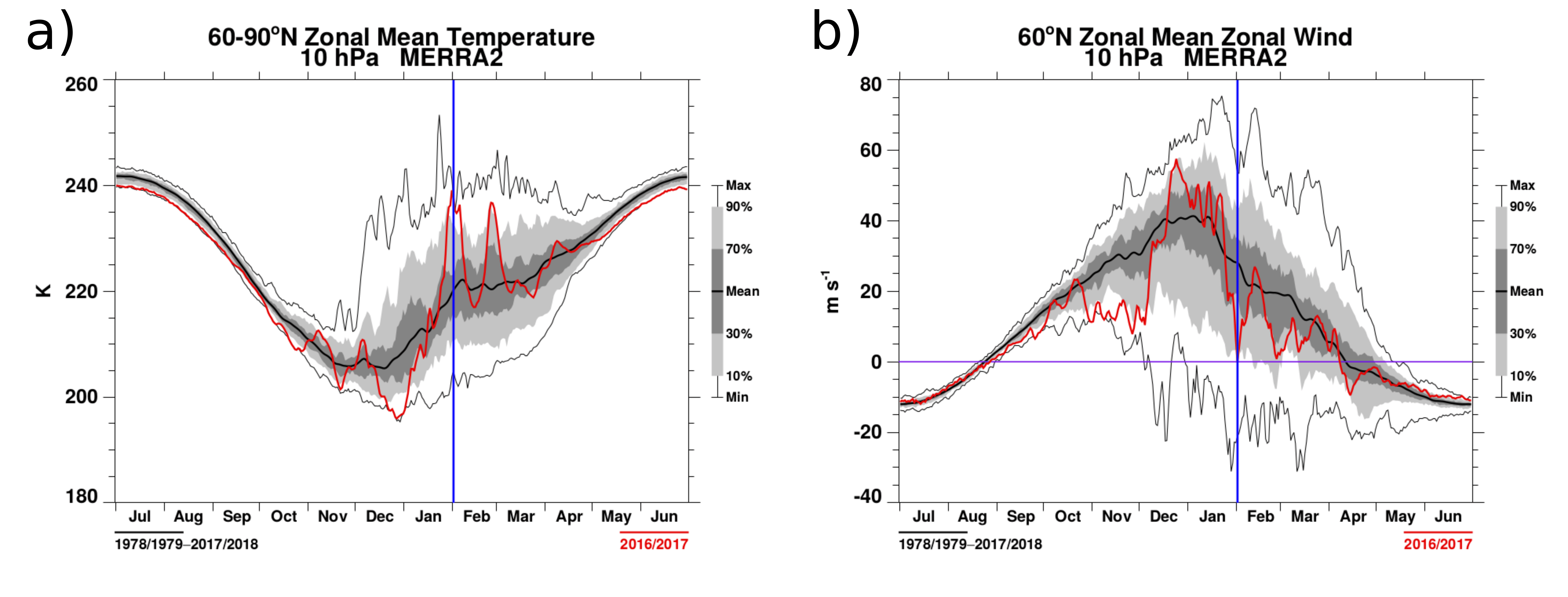}
	\caption{GEOS DAS MERRA-2 data used to define SSW events. a) zonal mean temperatures averaged over $60^{\circ}$N-$90^{\circ}$N. b) zonal mean zonal wind at $60^{\circ}$N. The red curve denotes values for the 2016-2017 period and the thick black curve corresponds to climatological values averaged from 1978 to 2018. The vertical
	blue lines reference a major SSW for that winter.}
	\label{fig:algorithm}
\end{figure}

Taking a closer look at Fig. \ref{fig:dfm}, we can see that an
anomalous increase in the effective temperature occurs between January and February 2017. 
The same anomalous behavior can be observed in the muon rate (see inset in Fig. \ref{fig:dfm}).
We used the \cite{charlton2007} definition and the Modern-Era Retrospective analysis for Research and Applications, Version 2 (MERRA-2), produced by the Goddard Earth Observing System Data Assimilation System (GEOS DAS) \citep{gelaro2017modern} to determine if a major SSW occurred during this time period.
We found that a major SSW took place during winter 2016-2017, with 
February 1 as the central date of the warming. 
In a few days, it increased the zonal mean temperature in the polar region by more than 20 K
(Fig. \ref{fig:algorithm} a).

\begin{figure}[h]
	\centering
	\includegraphics[width=1\linewidth]{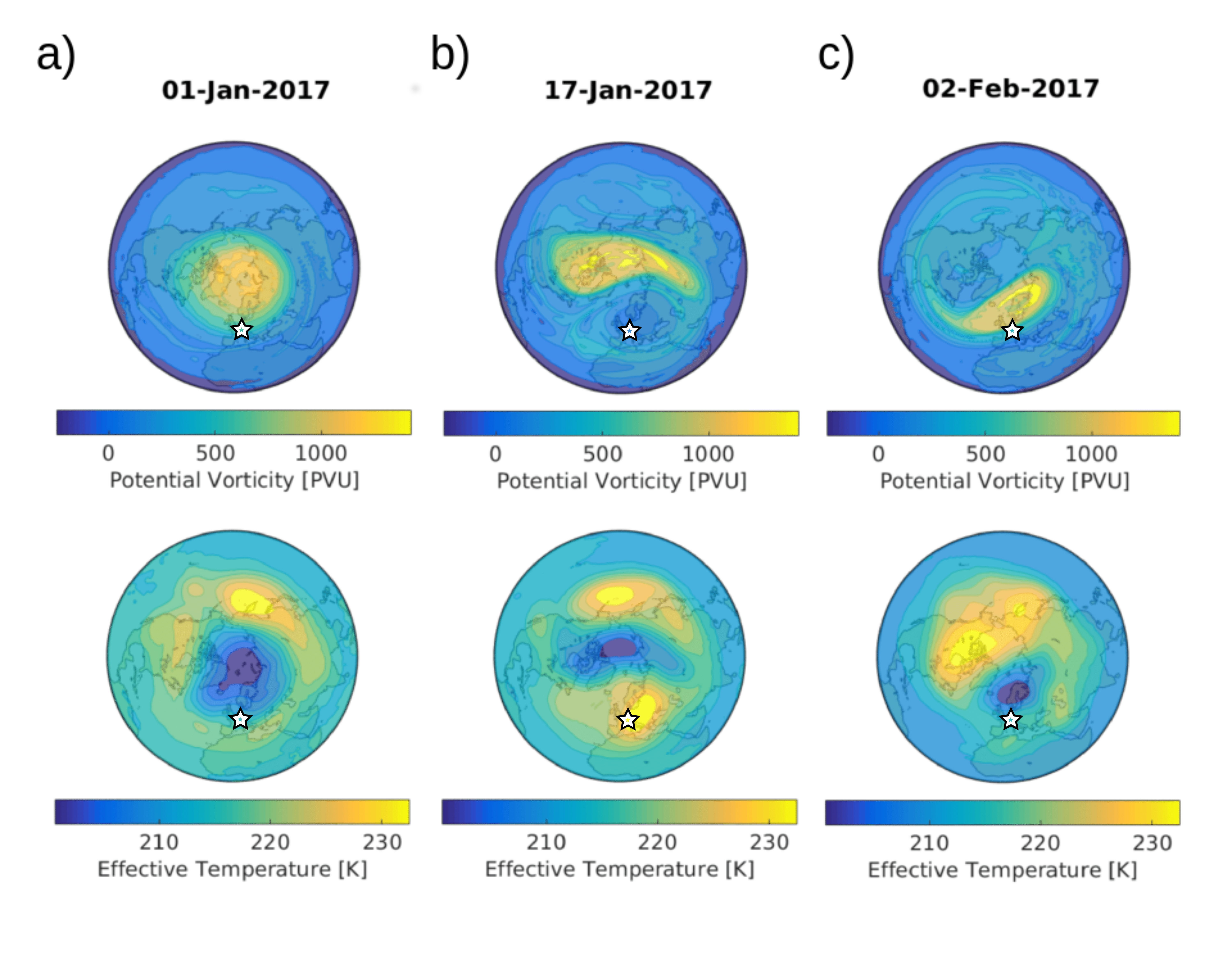}
	\caption{Potential vorticity at the 850 K potential temperature surface (top) and effective temperature
		(bottom) for January 1, January 17 and February 2, 2017, derived
		from the ECMWF data set. The maps are centered on the
                North Pole and the location of the Mont Terri
		Underground Laboratory ($47.38^{\circ}$N, $7.17^{\circ}$E),
		close to the town of Saint-Ursanne, Switzerland, is represented
		with a star. 1 PVU = $10^{-6} \ \text{K} \
		\text{m}^2 \ \text{Kg}^{-1} \ \text{s}^{-1}$.}
	\label{fig:ssw}
\end{figure}

Finally, we analyzed changes produced by the SSW using Ertel's
potential vorticity \citep{matthewman2009new}. This parameter quantifies the location, size, and shape of the
winter polar vortex. 
Figure \ref{fig:ssw} shows the spatial
distribution of Ertel's
potential vorticity at the 850 K potential temperature surface ($\sim 10$ hPa, $\sim 32$ km) for 3 different days, which are representative of the changes provoked.
The figure also shows the effective temperature spatial distribution during 
these 3 days. On January 1 (Fig. \ref{fig:ssw} a)
the vorticity and temperature exhibit ``typical'' winter conditions: the
polar vortex is centered on the Pole, together with the minimum
effective temperature. On January 17, a reshaping on the polar
vortex can be already observed. It is at this moment also that the
largest effective temperature anomaly occurs in the Mont Terri region
(Fig. \ref{fig:ssw} b). On February 2, that is, one day after the
event can be properly classified as a major SSW due to the reversal of the
zonal mean zonal wind (see Fig. \ref{fig:algorithm} b), the polar vortex
shape is still anomalous with the ``comma'' shaped maximum of potential
vorticity now closer to the Mont Terri URL
(Fig. \ref{fig:ssw} c). 
At the same time, the effective temperature in the Mont Terri region
has decreased to values similar to those in January 1.

\section{Discussion}

After a year of continous muon measurements with a portable muon detector under relatively low-opacity
conditions, we found that changes in the thermal
state of the atmosphere represent the largest cause of muon rate
variations. The correlation between these variables was first suggested
by a simple comparison of the relative
variation time-series. Then, it was confirmed by the large correlation
coefficient (0.81), and by the fitted effective temperature
coefficient, which is in agreement with the theoretical value
predicted for our particular opacity and zenith angle
conditions. Furthermore, our experiment was by chance performed under
similar opacity conditions to the Daya Bay detector, an established underground muon detector especially built for
neutrino experiments \citep{an2018seasonal}. Its corresponding
estimate of the effective temperature coefficient is also in agreement
with ours (Fig. \ref{fig:expValues}). 

Our muon detector is sensitive to both seasonal and short-term temperature variations.
The regional thermal anomaly reaching its maximum around January 17,
2017 (Fig. \ref{fig:dfm}), is
coincident
with the polar vortex changing its shape from
a normal pole-centered circle to a displaced ``comma shaped" one (Fig.
\ref{fig:ssw}). This is a typical feature of a SSW
\citep{ONeill2003encyclopedia}. Furthermore, the criteria by
\cite{charlton2007} for declaring a
major SSW is accomplished 15 days later. The time difference can be
potentially explained by the zonally-averaged wind criteria used to
define major SSWs, against the local character of the temperature variations
affecting the production of high-energy muons. 

Under much higher
opacity conditions (3,800 in mwe, i.e., more than 5 times the Mont
Terri URL opacity), the large muon detector of the Borexino experiment, 
Gran Sasso, Italy, also
reported muon rate variations related to this
SSW in 2017 \citep{agostini2019modulations}. 
Given the large opacity, most of the muons completely loose
their energy before reaching the detector. 
Thus, only high-energy muons resulting
from the decay of high-energy parent mesons are detected.
As explained by \cite{grashorn2010atmospheric}, high-energy
mesons are most sensitive
to middle-atmosphere temperature
variations due to their relatively longer lifetime, and thus a higher
probability of interacting with the atmosphere before decaying. 
This results in a higher sensitivity to temperature variations, which
translates into a larger effective temperature coefficient
(see Fig. \ref{fig:expValues}). Despite being in less advantageous
conditions in terms of detector acceptance and tunnel depth, our
portable muon detector was also able to detect these
short-term effect (15-days) directly linked to
middle-atmosphere dynamics (Fig. \ref{fig:dfm}). 

Compared to lidar measurements, 
which can obtain temperature profiles over tens of kilometers in altitude but have very narrow global coverage (only as wide as the laser beam),
muon detectors naturally provide integrated measurements in altitude, and a larger horizontal coverage.
Our results therefore imply that small and affordable muon detectors could
be used to study middle-atmosphere temperature variations without
resorting to, for example, expensive lidar systems. Besides being transportable, the
advantage is that no high-opacity conditions are needed. A
minimum opacity of 50 mwe would be required
to filter out the temperature-dependent lowest-energy muons
\citep{grashorn2010atmospheric}.
Besides being temperature dependent, low-energy muons can also be influenced by other phenomena such as atmospheric pressure variations \citep{jourde2016monitoring}, which is why we consider optimal to remove them. However, open-sky conditions may also reveal new insights into atmospheric phenomena (e.g., \cite{hariharan2019measurement}) and more experimental studies are needed to better understand the limits of the methodology.
Thus, detectors could be installed in any buried facility with access to
electrical power and
real-time data transmission, for example with a wi-fi network., such as road tunnels. 
In Europe, many underground research
facilities exist in this condition (e.g. Mont Terri UL in
Switzerland, 47.4$^{\circ}$N; the LSBB UL in France, 43.9$^{\circ}$N;
Canfranc UL in Spain,
42.7$^{\circ}$N). 
These experiments could be crucial
to fill the current data gap related to middle-atmospheric
dynamics,
in particular the study of temperature anomalies
associated to SSW in mid-latitudes
\citep{sox2016connection}.
Furthermore, the technique may be used to study
similar phenomena in the Southern Hemisphere.

The effective atmospheric temperature to which the muon rate is
 sensitive is a weighted average of a temperature profile from 0 to
50 km, with increasingly significant weights at higher altitudes
\citep{grashorn2010atmospheric}. Indeed, 70 $\%$ of the total weights
are given between 50 and 26 km, 90 $\%$  between 50 and 18 km and 95 $\%$ 
between 50 and 15 km (see Fig. \ref{fig:TandW}). 
Thus, muon rate variations are mostly sensitive to temperature
variations in the high stratosphere.
Muon measurements can therefore complement lidar mesospheric studies (e.g., \citet{sox2016connection,yuan2012wind}).
In terms of the spatial support, in the configuration used for this experiment (see Section 2), the total angular aperture of the detector is of approximately $\pm 40^{\circ}$, but more than 95\% of the muons are registered within an aperture of $\pm 30^{\circ}$. At 50 km, this represents a surface of 50$\times 50$ km${}^2$. Therefore, muon measurements may be used to sample more regional atmospheric behavior.

Besides the potential applications to atmospheric studies, 
portable muon
detectors may be used to precisely calibrate the effective temperature
curve (Fig. \ref{fig:expValues}). The experimental setups
used to estimate these values, so far, are concentrated in either high or low-opacity conditions, whereas with our approach we could sample 
the curve rather uniformly, even in the same tunnel by varying the orientation of our
detector and thus the opacity and zenith angle conditions. 

Our findings have direct implications for
applications aiming to characterize density variations in the
subsurface  (e.g. \citet{jourde2016muon}). 
Indeed, synchronous tracking of the open-sky muon
rate while performing a continuous imaging of a geological body (e.g. density monitoring) 
may not be sufficient to characterize the influence of high-atmosphere temperature
variations since the
relative effect on the total amount of muons registered increases with
opacity. In turn, the mentioned possibility to
improve the calibration of the muon-rate dependence with 
middle-atmosphere dynamics will be crucial to safely remove this
effect. The effect will be increasingly important at higher
latitudes due to the increase of seasonal
temperature variations, and for increasing rock opacities. 
At Mont Terri ($47.38^{\circ}$N), relative effective temperature variations can
be as high as 4$\%$, which given the effective temperature coefficient
estimated, imply changes in muon rate as high as $3\%$ (c.f. Fig. \ref{fig:dfm}). However, muon rate changes would be at
maximum $1\%$ if the opacity would be reduced by one
order of magnitude to 70 mwe, or equivalently 26 m of standard rock,
and for vertical observations.

Finally, relative temperature and muon
rate variations are not always coincident in Fig. \ref{fig:dfm},
despite using the same time-averaging window. Equivalently, 
deviations from the linear relationship up to 2\% and mostly around 1\% can be
observed in Fig. \ref{fig:alpha}.
The deviations from a perfect correspondence are presumably due to physical phenomena influencing the muon rate other than the effective atmospheric temperature. Variations arising from changes in the primary cosmic rays, or changes in the geomagnetic field induced by solar wind typically have temporal scales that are much smaller (e.g. seconds to hours) or much larger (e.g. a solar cycle of $\sim$11 years).
Changes reported recently as induced by lower altitude atmospheric phenomena such as thunderclouds only lasted 10 minutes \citep{hariharan2019measurement}, and the low-energy muons affected by atmospheric pressure variations \citep{jourde2016monitoring} get filtered in the first meters of rock in our experiment. A much more likely explanation may be given by 
changes in the groundwater content of the
rock overlying the Mont Terri URL and will be the subject of
forthcoming publications. 

\section{Conclusion}

We report for the first time sensitivity to middle-atmosphere
temperature variations using a portable muon detector. Changes
detected are associated not only to seasonal variations but also
short-term  (15-days) variations caused by a Sudden Stratospheric
Warming.
The occurrence of this event was verified by applying a standard definition of SSWs,
and also observed by regional temperature and polar vortex
variations obtained from ECMWF and MERRA-2 reanalysis data. 
Previous reports on the sensitivity of muon rate to these phenomena
exist only for large, expensive and immobile muon detectors
often times associated to neutrino experiments and high-opacity
conditions. 
Our findings imply that portable muon detectors may be used to further
study short-term temperature variations, and to improve the
calibration curve of muon rate dependence with an effective
temperature value. This, in turn, is crucial for
geoscience applications aiming at studying subsurface processes by
characterizing density changes with muons.

\acknowledgments
This study is part of the DIAPHANE project  and
was financially supported by the  ANR-14-CE 04-0001
and the MD experiment of the Mont Terri project (\url{www.mont-terri.ch})
funded by Swisstopo. MRC thanks the AXA Research Fund for
their financial support. We are grateful to Thierry Theurillat and
Senecio Schefer for their
technical and logistical assistance at Mont Terri URL.
The MERRA data are available from
\url{https://acd-ext.gsfc.nasa.gov/Data_services/met/ann_data.html}, and the
ECMWF data from \url{https://www.ecmwf.int/}.
Muon data used for all calculations are displayed in figures and are
available in the Supplementary Table S1.
This is IPGP contribution number 4049. 
We thank the editor and two anonymous reviewers for their constructive comments and suggestions, which helped to improve our work.


%

%




\end{document}